\documentclass[twocolumn,aps,prl,floatfix,superscriptaddress]{revtex4-1}
\usepackage{amsmath,amssymb}
\usepackage{graphicx,float}
\usepackage{times,txfonts}
\usepackage{color}
\usepackage{multirow}
\usepackage{bbold}
\usepackage{color}
\usepackage{soul}
\usepackage{hyperref}
\usepackage{times,txfonts}
\usepackage{natbib}
\usepackage{blindtext}
\usepackage[export]{adjustbox}
\usepackage{mathrsfs}
\usepackage{textcomp}
\usepackage{blkarray}
\usepackage{multirow}

\newcommand{\ket}[1]{\left| #1 \right\rangle}

\renewcommand{\epsilon}{\varepsilon}
\renewcommand{\phi}{\varphi}

\def\VR{\kern-\arraycolsep\strut\vrule &\kern-\arraycolsep}
\def\vr{\kern-\arraycolsep & \kern-\arraycolsep}
\usepackage{sistyle}
\usepackage{soul}
\definecolor{lightblue}{RGB}{185,210,248}
\sethlcolor{lightblue}

\begin{document}
\title{Underwater Quantum Key Distribution in Outdoor Conditions with Twisted Photons}
\author{Fr\'ed\'eric Bouchard}
\email{fbouc052@uottawa.ca}
\affiliation{Physics Department, Centre for Research in Photonics, University of Ottawa, Advanced Research Complex, 25 Templeton, Ottawa ON Canada, K1N 6N5}
\author{Alicia Sit}
\affiliation{Physics Department, Centre for Research in Photonics, University of Ottawa, Advanced Research Complex, 25 Templeton, Ottawa ON Canada, K1N 6N5}
\author{Felix Hufnagel}
\affiliation{Physics Department, Centre for Research in Photonics, University of Ottawa, Advanced Research Complex, 25 Templeton, Ottawa ON Canada, K1N 6N5}
\author{Aazad Abbas}
\affiliation{Physics Department, Centre for Research in Photonics, University of Ottawa, Advanced Research Complex, 25 Templeton, Ottawa ON Canada, K1N 6N5}
\author{Yingwen Zhang}
\affiliation{Physics Department, Centre for Research in Photonics, University of Ottawa, Advanced Research Complex, 25 Templeton, Ottawa ON Canada, K1N 6N5}
\author{Khabat Heshami}
\affiliation{National Research Council of Canada, 100 Sussex Drive, Ottawa ON Canada, K1A 0R6}
\author{Robert Fickler}
\affiliation{Physics Department, Centre for Research in Photonics, University of Ottawa, Advanced Research Complex, 25 Templeton, Ottawa ON Canada, K1N 6N5}
\author{Christoph Marquardt}
\affiliation{Max-Planck-Institut für die Physik des Lichts, Staudtstraße 2, 91058 Erlangen, Germany}
\affiliation{Institut für Optik, Information und Photonik, Universität Erlangen-Nürnberg, Staudtstraße 7/B2, 91058 Erlangen, Germany}
\author{Gerd Leuchs}
\affiliation{Physics Department, Centre for Research in Photonics, University of Ottawa, Advanced Research Complex, 25 Templeton, Ottawa ON Canada, K1N 6N5}
\affiliation{Max-Planck-Institut für die Physik des Lichts, Staudtstraße 2, 91058 Erlangen, Germany}
\affiliation{Institut für Optik, Information und Photonik, Universität Erlangen-Nürnberg, Staudtstraße 7/B2, 91058 Erlangen, Germany}
\author{Robert~W.~Boyd}
\affiliation{Physics Department, Centre for Research in Photonics, University of Ottawa, Advanced Research Complex, 25 Templeton, Ottawa ON Canada, K1N 6N5}
\affiliation{Max-Planck-Institut für die Physik des Lichts, Staudtstraße 2, 91058 Erlangen, Germany}
\affiliation{Institute of Optics, University of Rochester, Rochester, New York, 14627, USA}
\author{Ebrahim Karimi}
\email{ekarimi@uottawa.ca}
\affiliation{Physics Department, Centre for Research in Photonics, University of Ottawa, Advanced Research Complex, 25 Templeton, Ottawa ON Canada, K1N 6N5}
\affiliation{Max-Planck-Institut für die Physik des Lichts, Staudtstraße 2, 91058 Erlangen, Germany}
\affiliation{Department of Physics, Institute for Advanced Studies in Basic Sciences, 45137-66731 Zanjan, Iran.}
%



%
%
\begin{abstract}
Quantum communication has been successfully implemented in optical fibres and through free-space~\cite{muller:93,buttler:98,rarity:01}. Fibre systems, though capable of fast key rates and low quantum bit error rates (QBERs), are impractical in communicating with destinations without an established fibre link~\cite{valivarthi:16}. Free-space quantum channels can overcome such limitations and reach long distances with the advent of satellite-to-ground links~\cite{yin:17a,yin:17b,ren:17,Liao:18}. Shorter line-of-sight free-space links have also been realized for intra-city conditions~\cite{buttler:98,resch:05}. However, turbulence, resulting from local fluctuations in refractive index, becomes a major challenge by adding errors and losses~\cite{andrews:05}. Recently, an interest in investigating the possibility of underwater quantum channels has arisen, which could provide global secure communication channels among submersibles and boats~\cite{shi:14,zhou:15,ji:17}. Here, we investigate the effect of turbulence on an underwater quantum channel using twisted photons in outdoor conditions. We study the effect of turbulence on transmitted QBERs, and compare different QKD protocols in an underwater quantum channel showing the feasibility of high-dimensional encoding schemes. Our work may open the way for secure high-dimensional quantum communication between submersibles, and provides important input for potential submersibles-to-satellite quantum communication.
\end{abstract}

\maketitle

%
%
Quantum key distribution (QKD) allows two individuals, conventionally referred to as \textit{Alice} and \textit{Bob}, to communicate information in a secure and secret manner~\cite{bennett:84}. Since the proposal of the first protocol by Bennett and Brassard in 1984 (BB84)~\cite{bennett:84}, various protocols and methods, for example Ekert91~\cite{ekert1991quantum} and six-state~\cite{liang:03}, have been further proposed and experimentally investigated. Notably, one class of quantum cryptographic schemes, namely high-dimensional QKD protocols, makes use of \textit{qudits} rather than qubits, wherein the encoded quantum states belong to a higher-dimensional Hilbert space~\cite{bechmann:00,cerf:02}. Such schemes have many potential advantages: in the case of an error-free channel, more than one bit of information can be distributed per carrier. Moreover, they tolerate larger error-thresholds due to the difficulties that an eavesdropper \textit{Eve} has in getting information about the high-dimensional state~\cite{bouchard:17}. This may allow for the implementation of QKD links in noisy environments with high QBER.

Photons are the carriers of choice for quantum communication, possessing multiple degrees of freedom with which information can be encoded. Polarization \cite{bennett:84}, time-bins~\cite{islam:17}, and spatial modes \cite{groblacher:06} are the most prevalent encryption methods, with the last two being common methods for achieving high-dimensional protocols. One family of spatial modes with mature preparation and measurement techniques is the OAM of light, also referred to as twisted photons~\cite{molina:07,erhard:17}. These modes possess a helical wavefront given by $\exp(i\ell\phi)$, where $\ell$ is an integer and $\phi$ is the transverse azimuthal coordinate. The OAM states of photons is one realization of a Hilbert space with unbounded dimensionality. Since the modes form a complete orthonormal basis, these states can be used for high-dimensional QKD schemes~\cite{mirhosseini:15,vallone:14,sit:17}. In this Letter, we report the effect of water turbulence on OAM modes of light in an outdoor swimming pool, and study its effect in quantum cryptographic schemes, performing a high-dimensional BB84 protocol with twisted photons.

%
%
Since the underwater quantum channel is an outdoor link, uncontrolled turbulent conditions can be expected, as in the case of free-space links. Turbulence is observed in the form of beam distortions and beam wandering after propagating through a turbulent media. The effect of turbulence on the propagation of OAM modes through free-space air has been studied for various distances. In the Kolmogorov theory of turbulence in free-space, the turbulence is associated with a local variation in the refractive index due to temperature and pressure variations~\cite{kolmogorov:41}. However, temperature gradients in the atmosphere represent the main contribution to atmospheric turbulence. Water is an incompressible fluid and thus the main contribution to the optical turbulence is derived from local variations in temperatures. Recently, propagation of OAM modes through water has been reported in controlled laboratory conditions~\cite{ren:16,baghdady:16}. 

\begin{figure*}[!htbp]
	\begin{center}
	\includegraphics[width=2\columnwidth]{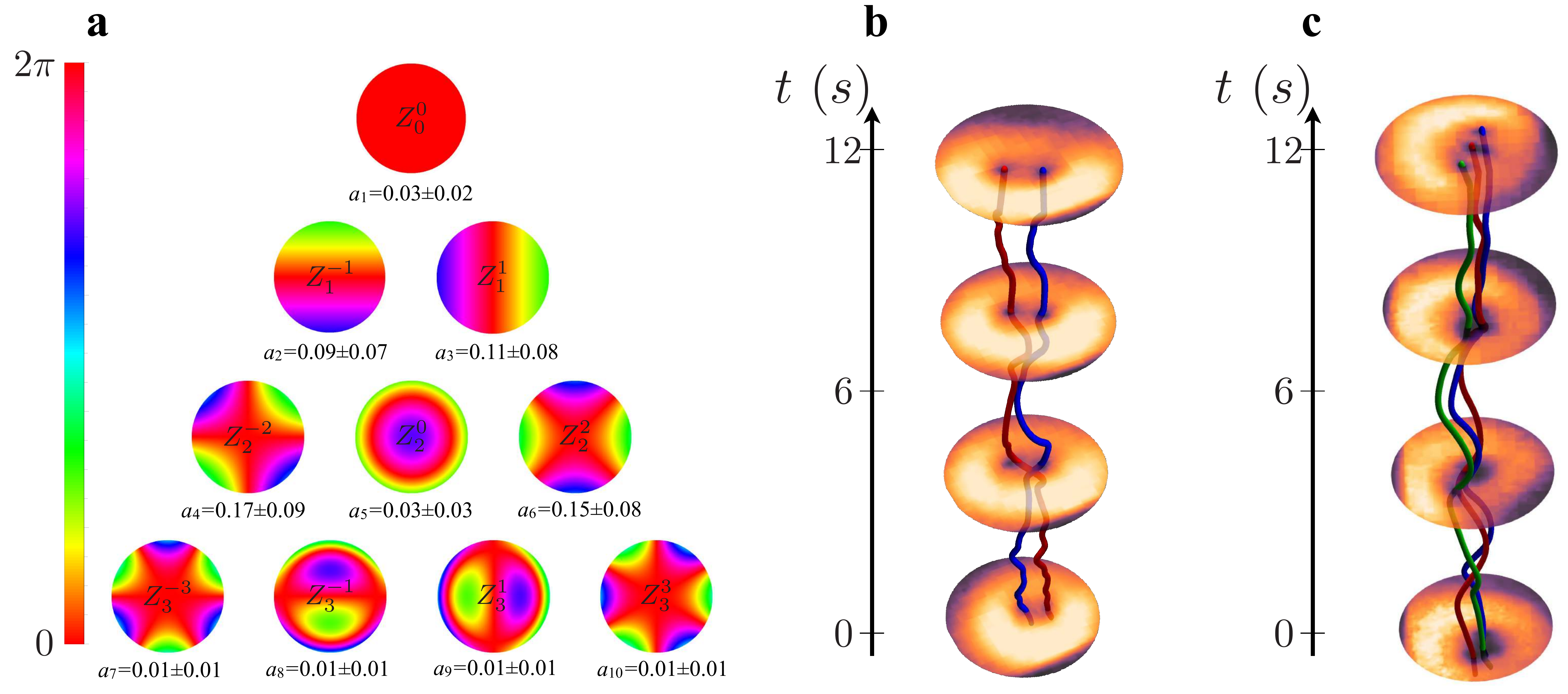}
	\caption[]{\textbf{Experimental characterization of underwater turbulence.} \textbf{a.} Calculated coefficients for the lowest ten Zernike polynomials from intensity images of a Gaussian beam after propagation through 3~m of water to characterize the turbulence in one particular set of conditions at the time of measurement. The dominant coefficients correspond to oblique and vertical astigmatism ($a_4$ and $a_6$), followed by tip and tilts effects ($a_2$ and $a_3$). \textbf{b} and \textbf{c.} Evolution of vortex splitting over a 12~s period for a $\ell=2$ and $\ell=3$ modes, respectively, sent through 5~m of water. The red, blue, and green lines represent the trajectories of the individual singularities, highlighting their splitting and wandering that occurs due to the turbulence.}
	\label{fig:turb}
	\end{center}
\end{figure*}

A characterization of the level of turbulence, assuming the single phase screen approximation, in our 3~m underwater channel is performed by sending a 635~nm Gaussian-shaped laser beam through the water and record the transmitted intensity patters (see Turbulence Characterization in Methods for more details). We employ the Gerchberg-Saxton algorithm (GSA), a phase retrieval algorithm using fast Fourier transforms~\cite{fienup:82}, to reconstruct the phase of the beam after propagating through the water. The obtained phase profile, $\Phi(r,\phi)$, is then decomposed in terms of Zernike polynomials, which forms a set of orthonormal polynomials on the unit disk~\cite{noll:76}, 
\begin{eqnarray}
	\Phi (r,\phi) = \sum_j a_j Z_j(r,\phi),
\end{eqnarray}
where $r$ and $\phi$ are the radial and azimuthal coordinates, respectively, $a_j$ are the Zernike coefficients, $Z_j (r,\phi) = Z_n^m (r,\phi) $ are the Zernike polynomials (defined in the Methods), $j=1+(n(n+2)+m)/2$ is the Noll index,  and $n$ and $m$ are the radial and azimuthal degree, respectively.

The average values of measured expansion coefficients $a_j$ as well as their corresponding Zernike polynomials are shown in Fig.~\ref{fig:turb}-\textbf{a}. In particular, low-order Zernike polynomials have specific meaning in terms of optical aberrations. First order aberrations,  $n=1$ ($j=2,3$), correspond to a tip-tilt in the wavefront. In the weak atmospheric turbulence regime, tip-tilt is the major contribution and results in beam wandering. Second order optical aberrations, $n=2$, are related to astigmatism ($j=4,6$) and defocusing ($j=5$). It can be seen from Fig.~\ref{fig:turb}-\textbf{a}, that the contribution of astigmatism in our turbulent underwater link is the largest. In particular, one effect of astigmatism on OAM modes is the singularity splitting for OAM values of $|\ell|>1$; this splitting effect has also recently been studied in free-space~\cite{Lavery:17}. The effect of vortex splitting in our underwater link is shown in Fig.~\ref{fig:turb}-\textbf{b} and Fig.~\ref{fig:turb}-\textbf{c}, where an $\ell=2$ and $\ell=3$ mode respectively, each generated by a phase-only spatial light modulator (SLM), is sent through a slightly longer distance of 5~m. Hence, underwater channels may give rise to turbulent conditions that are fundamentally different from those present in a free-space channel. However, the turbulence was observed to change on a much slower time-scale as opposed to free-space, on the order of 10~Hz compared to 100~Hz. Thus, implementing a SLM in an adaptive optics type system might be fast enough to correct for the aberrations.

\begin{figure*}[!htbp]
	\begin{center}
	\includegraphics[width=1.8\columnwidth]{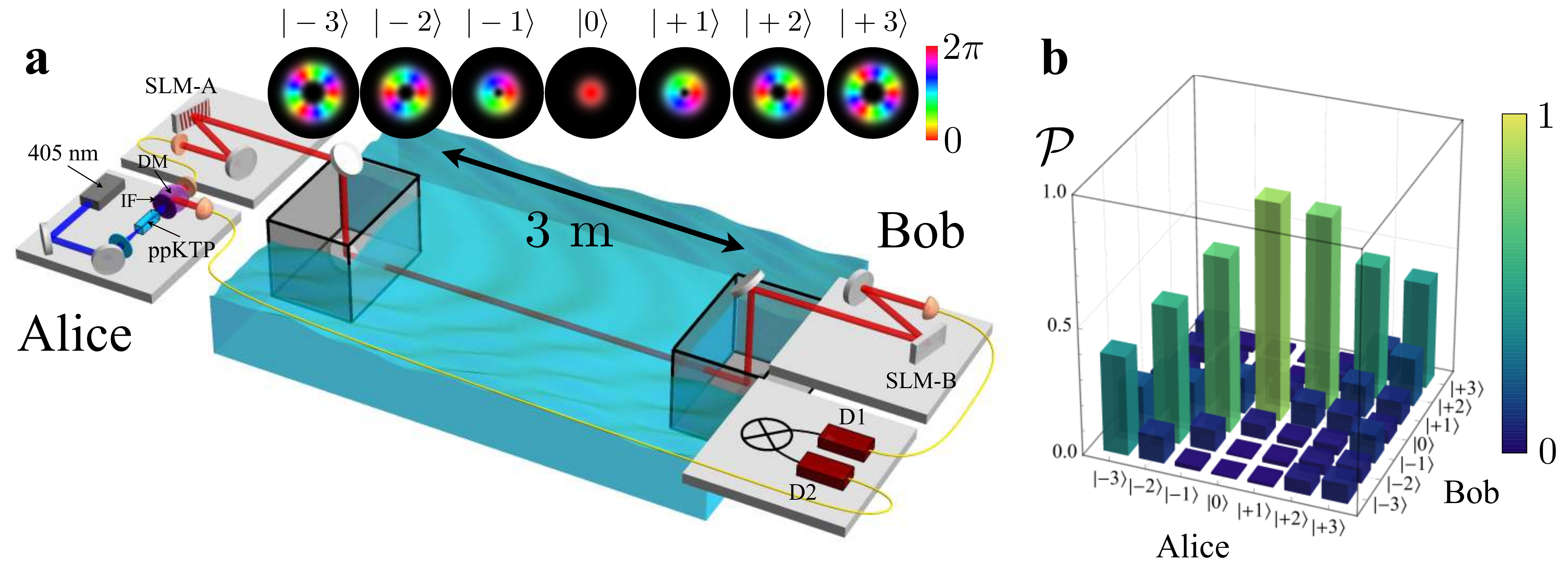}
	\caption[]{\textbf{Experimental setup and state cross-talk measurements.} \textbf{a.} Photon pairs (signal $\lambda_s=710~nm$, idler $\lambda_i=943~nm$) are generated via spontaneous parametric downconversion pumped from a periodically poled KTP (ppKTP) crystal by a 405~nm diode laser. A long pass filter (IF) blocks the UV and transmits the photon pairs, which are then split at a dichroic mirror (DM). The idler photon is directly detected by a single photon detector (D2) and acts a heralding trigger for the information-carrying signal photon. Alice prepares the signal photon into a particular state, for example one from the insets, using SLM-A, then sends it to Bob through the 3~m underwater link. Bob performs a measurement on the received state using SLM-B and a single mode optical fibre connected to D1. Coincidence events between D1 and D2 are recorded. \textbf{b.} Measured cross-talk matrix between the OAM states ($\ell=-3$ to 3) that Alice sends and Bob measures. Higher order states experience more cross-talk as compared to lower order states, seen as off-diagonal detection probabilities.}
	\label{fig:exp}
	\end{center}
\end{figure*}

%
%
Our experimental setup for investigating QKD consists of a heralded single photon source (for more details see Experimental Setup in Methods), Alice's state preparation setup, Bob's measurement setup, and a 3~m-outdoor underwater link --- an outdoor pool with uncontrolled conditions --- see Fig.~\ref{fig:exp}-\textbf{a}.  In the near-infrared region, light is strongly absorbed by water; ideally, it is desirable to produce signal photons with a $\lambda_s$ in the blue-green window ($\approx$400-600~nm) which experiences the least amount of absorption. In the heralded single-photon source, the signal ($\lambda_s$ = 710~nm) and idler ($\lambda_i$ = 940~nm) photons are generated by spontaneous parametric downconvesion, and are coupled to single-mode optical fibres (SMOF) in order to filter their transverse spatial modes to the fundamental Gaussian mode. A coincidence rate of $432~\mathrm{kHz}$, within a coincidence time window of 5~ns, is measured after the SMOFs at the source. The idler photon is sent through a fibre delay line to Bob, acting as the heralding photon, and the signal photon is sent to Alice's generation apparatus. In order to eliminate the distortions that an air-water interface would introduce to the wavefront of the transmitted and recieved photons, we use periscopes to guide the photons into/out of glass tanks that are partially immersed in the water on either end of the link. The advantage of using such a configuration is that the photons pass through first a flat air-glass then a glass-water interface, and \textit{vice versa}, without significant alterations to their wavefronts. For the quantum cryptographic tests, Alice prepares the signal photon into an OAM state using a SLM, then sends it across the underwater link. Bob uses a SLM and SMOF to project the received signal photons onto a given OAM state and records a coincidence event between the result and the heralding photon at a coincidence box~\cite{qassim:14}.

%
%
We perform a cross-talk measurement of several OAM states ranging from $-3$ to $3$, i.e. $\{ \ket{\ell}; \ell=-3,-2,-1,0,1,2,3 \}$, see Fig.~\ref{fig:exp}-\textbf{b}, where $\ket{\ell}$ represents the quantum state with helical wavefront of exp($i\ell\phi$). The cross-talk measurements are a good indicator of the level of errors (QBER) that one could expect in a QKD protocol. Practical implementations are seen to dictate the optimal dimensionality of the qudits used in a specific high-dimensional quantum cryptographic scheme. The OAM mode that experiences the least amount of cross-talk is the fundamental Gaussian mode ($\ell=0$), with a cross-talk of $<15\%$ with its neighbouring modes ($\ell=\pm1$). This cross-talk could lead to sufficiently low QBER to securely transmit information, given a small OAM encryption subspace. As we go to larger OAM values, the modes suffer larger cross-talk, which makes the extension to higher-dimensions challenging. Explicitly, the effect of turbulence on a QKD protocol is twofold: it introduces errors and losses. Most QKD protocols are robust against losses at the cost of a reduced key rate. However, the effect of errors is more critical since the protocol must be aborted if the error level exceeds a set threshold. 


\begin{figure*}[!htbp]
	\begin{center}
	\includegraphics[width=1.8\columnwidth]{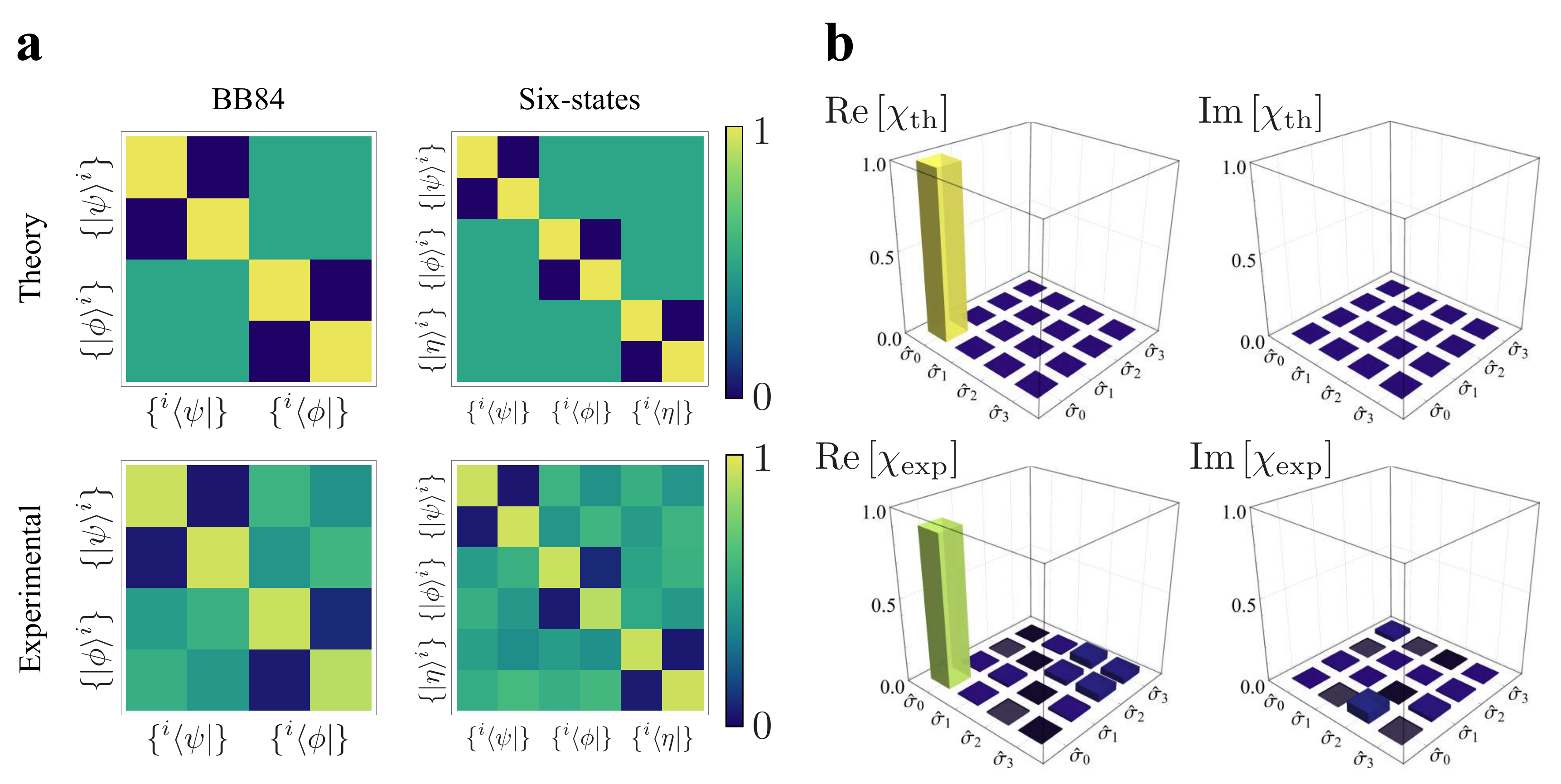}
	\caption[]{\textbf{Probability-of-detection matrices for $d$=2 BB84 and Six-states protocols, and the channel process matrix.} \textbf{a.} Theoretical and experimentally measured probability-of-detection matrices for BB84 (left column) and Six-states (right column) protocols in $d=2$. We measured QBERs of $Q=6.57\%$ and $Q=6.35~\%$, respectively, for these two protocols, corresponding to secret key rates of $R=0.301$ and $R=0.395$. \textbf{b.} The six-state protocol is a tomographic protocol and can be used to reconstruct the process tomography matrix; the real and imaginary parts of the theoretical matrix are shown in the top row. The experimentally measured process matrix is shown in the bottom row with a process fidelity of ${\cal F} = 0.905$.}
	\label{fig:2d}
	\end{center}
\end{figure*}

\begin{figure}[!htbp]
	\begin{center}
	\includegraphics[width=1.0\columnwidth]{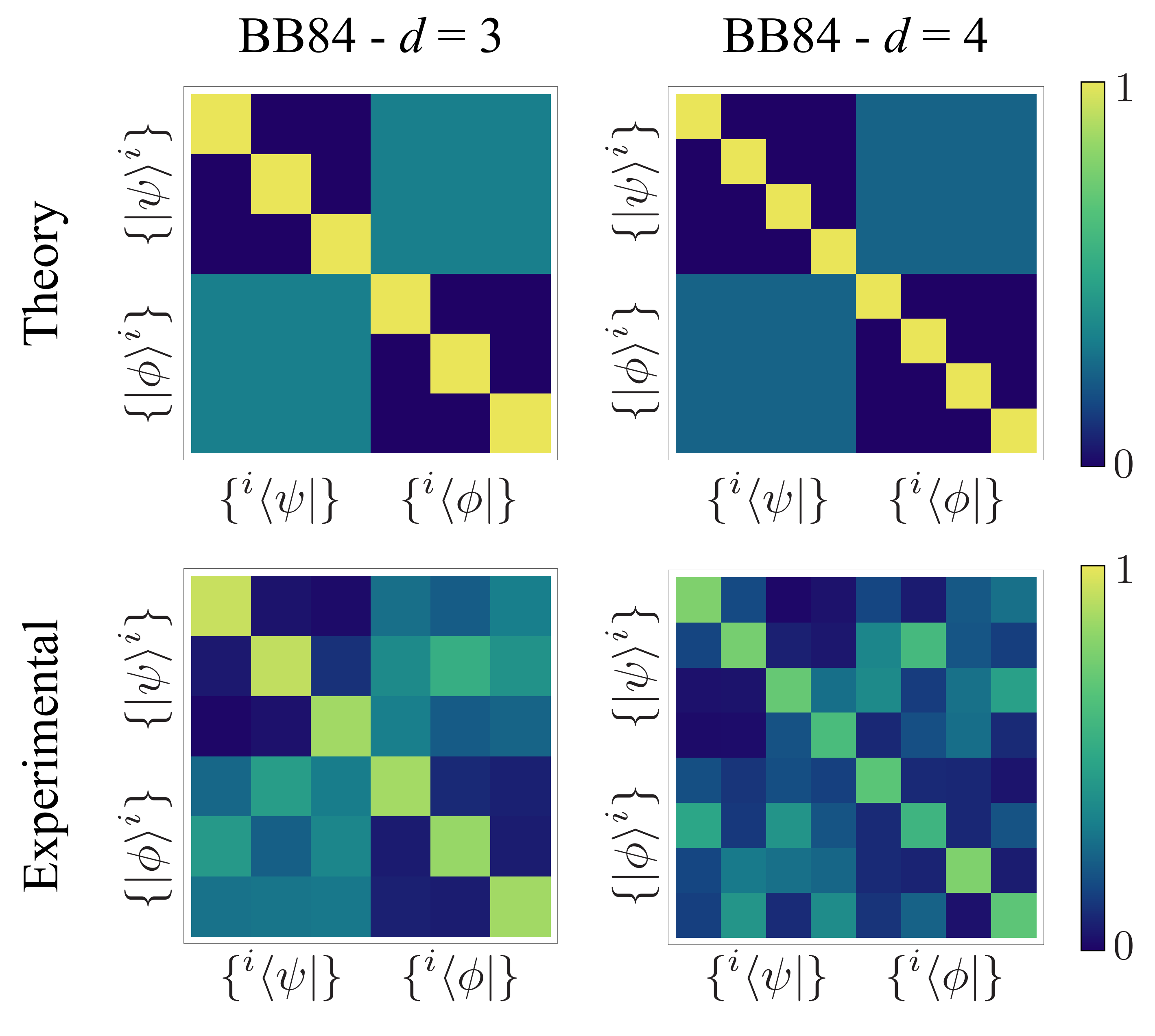}
	\caption[]{\textbf{High-dimensional probability-of-detection matrices.} Theoretical (top row) and experimentally measured (bottom row) probability-of-detection matrices for BB84 protocols in $d=3$ and $d=4$. We measured QBER of $Q^\mathrm{3D}=11.73\%$ and $Q^\mathrm{4D}=29.77\%$, respectively. The QBER in $d=3$ is below the tolerable error threshold, allowing for the establishment of a secret key rate of $R^{3D}=0.307$ bits per sifted photon. However, the QBER in $d=4$ exceeds the threshold of $Q^\mathrm{4D}_\mathrm{threshold}=18.9\%$. }
	\label{fig:hd}
	\end{center}
\end{figure}

As a first test of our underwater QKD link, we perform a 2-dimensional BB84 protocol. Alice uses the OAM subspace consisting of $\ell=\pm1$ to encode the information. In the BB84 protocol, two mutually unbiased bases (MUBs) are required for Alice and Bob to encode and measure the states of the photons. The first MUB here is given by the logical basis, $\ket{\psi}^i \in \{\ket{-1} , \ket{+1}\}$, and the second MUB is given by $\ket{\phi}^i \in \{ \ket{+}, \ket{-} \}$, where $\ket{\pm} = (\ket{-1} \pm \ket{+1})/\sqrt{2}$. The experimental probability-of-detection matrix is shown in Fig.~\ref{fig:2d}-\textbf{a} (left column) along with its theoretical counterpart. The secret key rate per sifted photon, $R$, may be calculated using the following formula, $R(Q) = 1 - 2 h(Q)$, where $Q$ is the QBER and $h(\cdot)$ is the Shannon entropy.  From the probability-of-detection matrix, a QBER of $Q=6.57~\%$ is calculated, which is below the error threshold of $Q^\mathrm{2D}_\mathrm{threshold}=11\%$ for the 2-dimensional BB84 protocol, corresponding to a positive secret key rate of $R=0.301$ bits per sifted photon. 

An extension of the BB84 protocol in dimension $d=2$ is achieved by considering a third MUB, i.e. $\ket{\eta}^i \in \{ \ket{+i}, \ket{-i} \}$, where $\ket{\pm i} = (\ket{-1} \pm i \ket{+1})/\sqrt{2}$. This protocol, also known as the \textit{Six-states} protocol~\cite{bruss:98}, can tolerate slightly larger error thresholds of around $Q=12.6~\%$. The probability-of-detection matrix is shown in Fig.~\ref{fig:2d}-\textbf{a} (right column), where a QBER of $Q=6.35~\%$ is measured resulting in a secret key rate of $R=0.395$ bits per sifted photon.  However, when considering sifting, the \textit{six-states} protocol suffers from a lower sifting rate, i.e. 1/3, in comparison to the BB84 protocol, which has a sifting rate of 1/2. Nevertheless, the \textit{six-states} protocol is a tomographic protocol: the measurements by Alice and Bob can be used to fully characterize the quantum channel and reconstruct the process matrix of the link via quantum process tomography. Let the channel be characterized by a process $\varepsilon$, which relates the input and output states in the following manner, $\hat{\rho}_\mathrm{out} = \varepsilon \left( \hat{\rho}_\mathrm{in} \right)$. The process may be described by the process matrix $\chi_{mn}$, where $\varepsilon \left( \hat{\rho} \right) = \sum_{mn} \chi_{mn} \ \hat{\sigma}_m \  \hat{\rho} \  \hat{\sigma}_n^\dagger $, and $\hat{\sigma}_m$ are the Pauli matrices. The reconstructed process matrix, $\chi_{\text{exp}}$, along with the theoretical ideal process matrix, $\chi_{\text{th}}$, is shown in Fig.~\ref{fig:2d}-\textbf{b}. A process fidelity of ${\cal F} = 0.905$ is measured from the process matrix, where the process fidelity is defined as ${\cal F}=\mathrm{Tr} \left[ \chi_{\text{exp}} \cdot \chi_{\text{th}} \right]/ \mathrm{Tr} \left[ \chi_{\text{th}} \cdot \chi_{\text{th}} \right]$.

The versatility of our experimental configuration allows us to test different types of QKD protocols in our underwater link. As a next step, we perform a high-dimensional quantum cryptographic scheme. The standard BB84 protocol is naturally extended using high-dimensional states, where two $d$-dimensional bases are employed. The first MUB is given by the logical basis, $\ket{\psi}^i \in \{\ket{i}; i=1,2, ... ,d\}$, and the second MUB is given by the discrete Fourier transform $\ket{\phi}^i \in \{\frac{1}{\sqrt{d}} \sum_{j=0}^{d-1} \omega_d^{ij} \ket{i}  \}$, where $\omega_d = \exp (i 2 \pi /d)$. We perform the 3- and 4-dimensional BB84 protocol using the OAM modes with $\ell=0,\pm1$ and $\ell=\pm1,\pm2$, respectively, in our underwater link. The results are shown in Fig.~\ref{fig:hd}, where QBERs of $Q^\mathrm{3D}=11.73\%$ and $Q^\mathrm{4D}=29.77\%$ were measured for the case of $d=3$ and $d=4$, respectively. For the 3-dimensional BB84 ($Q^\mathrm{3D}_\mathrm{threshold}=15.95\%$), a secret key rate of $R^\mathrm{3D}=0.307$ bits per sifted photon was obtained, which is slightly larger than the 2-dimensional BB84 secret key rate. For the 4-dimensional case, the QBER is above the error threshold, i.e. $Q^\mathrm{4D}_\mathrm{threshold}=18.93\%$, meaning no secret key can distributed across the turbulent underwater link with a 4-dimensional BB84 protocol with twisted photons. These errors originate from the aberrations induced by the underwater turbulence, introducing more cross-talk between higher OAM states. As mentioned previously, the frequency of the turbulence was on the order of tens of Hertz, which opens up the possibility to implement an adaptive optics system using the implemented SLMs on Alice's or Bob's side for correcting the aberrations. This procedure would provide a means for reducing the QBER below the error thresholds in higher-dimensions.

%
%
In summary, we have characterized the predominant turbulence effects in our underwater quantum channel to be astigmatism, outlining a notable difference between an air free-space and an underwater link. We have performed and compared different QKD protocols through this underwater link using twisted photons. For a short distance, i.e. 3~m, we were able to successfully achieve a positive secret key rate using a 2- and 3-dimensional BB84 protocol.
\\

%
%
\vspace{0.5cm}
\noindent\textbf{Methods}\newline

\noindent{\footnotesize{\bf Turbulence Characterization:} A characterization of the level of turbulence in our underwater channel is done by sending a Gaussian laser beam, at a wavelength of 635~nm, over our 3~m underwater link. Short exposure images (0.07~ms) of the beam at the output of the link are recorded using a CCD camera. The water turbulence is characterized using a single phase screen approximation, i.e. we assumed the effect of turbulence can be described as a varying phase screen at the input of the link followed by uniform propagation. Assuming a Gaussian input beam, we use the intensity images recorded at the output of the link to reconstruct the phase of the input beam. The reconstructed input phase profile corresponds to the input single phase screen that models the turbulence of the channel. In order to obtain the phase of the output beam, we perform the Gerchberg-Saxton algorithm (GSA), a phase retrieval algorithm using fast Fourier transforms~\cite{fienup:82}. The obtained phase profile, $\Phi(r,\phi)$, is then decomposed in terms of Zernike polynomials, which forms a set of orthonormal polynomials on the unit disk, $\Phi (r,\phi) = \sum_j a_j Z_j(r,\phi)$ as defined in the main text. Explicitly, the Zernike polynomials are written in terms of the radial polynomial $R_n^m (r)$~\cite{noll:76},
\begin{eqnarray}
	Z_{\mathrm{even}\, j} \,(r, \phi) &=& \sqrt{n+1}\,R_n^m (r) \sqrt{2} \cos (m \phi ), \,\,\,\,\, m\neq0, \\
	Z_{\mathrm{odd}\, j} \,(r, \phi) &=& \sqrt{n+1}\,R_n^m (r) \sqrt{2} \sin (m \phi ), \,\,\,\,\,\, m\neq0, \\
	Z_{j} \,(r, \phi) &=& \sqrt{n+1} R_n^0 (r), \,\,\,\,\,\, m=0.
\end{eqnarray}
The GSA and Zernike polynomial decomposition is subsequently carried over all 143 images recorded at the output of the link. \newline

\noindent{\footnotesize{\bf Experimental Setup:} In the heralded single photon source, a 405~nm diode laser (200~mW) pumps a periodically-poled potassium titanyl phosphate (ppKTP) crystal to produce single photon pairs via spontaneous parametric downconversion. A non-degenerate set of wavelengths is chosen to produce signal photons at $\lambda_s=710$~nm, with corresponding idler photons at $\lambda_i = 943$~nm. We note that the wavelength of the signal photon could be adjusted to lie in the desired blue-green window with a different crystal along with commercially available single photon detectors which work at the IR.  The signal and idler photons are coupled to single-mode optical fibres (SMOF) in order to filter their transverse spatial modes to the fundamental Gaussian mode. A coincidence rate of $432~\mathrm{kHz}$, within a coincidence time window of 5~ns, is measured after the SMOFs at the source. The corresponding single photon count rates for the signal and idler photons are given by 5~MHz and 1.5~MHz, respectively. The idler photon is sent through a fibre delay line to Bob, acting as the heralding photon, and the signal photon is sent to Alice's generation apparatus. The experiment was carried out during the night under the following weather conditions: temperature, relative humidity, wind speed and atmospheric pressure were measured as $17^\circ$C, $91\%$, $2$~km/h and $100.79$~kPa, respectively. The depth of the pool is $1.1$~m and the beam was situated at $12$~cm under the surface. The pH, Phosphate concentration, and water hardness were measured as 6.9, 318 ppb and 331 ppm, respectively.

\bibliographystyle{naturemag}

%
%
\vspace{0.2cm}
\noindent\textbf{Author Contributions}
\noindent F.B., A.S, F.H. contributed equally to this work. F.B., A.S., F.H., A.A. performed the experiment. F.B., A.S., F.H., K.H., and R.F. analyzed the data. Y.Z., R.F., C.M., G.L., R.W.B. and E.K. supervised the project. F.B., A.S., F.H., and E.K. wrote the manuscript with the help of the other co-authors.
\vspace{0.5 EM}

\noindent\textbf{Acknowledgments} All authors would like to thank Norman Bouchard and Marie-France Langlois for access to their in-ground pool. This work was supported by Canada Research Chairs; Canada Foundation for Innovation (CFI); Canada Excellence Research Chairs, Government of Canada (CERC); Canada First Research Excellence Fund (CFREF); Natural Sciences and Engineering Research Council of Canada (NSERC); and Max Planck-University of Ottawa Centre for Extreme and Quantum Photonics.

\vspace{0.5 EM}

\noindent\textbf{Author Information}
\noindent Correspondence and requests for materials should be addressed to ekarimi@uottawa.ca.

\vspace{0.5 EM}

\noindent\textbf{Competing interests} The authors declare no competing financial interests.

\end{document}